\documentclass[aps,prb,amsmath,showpacs,preprint,superscriptaddress]{revtex4}
\usepackage{bm}
\usepackage{graphicx}
\DeclareMathOperator{\sgn}{sgn}

\DeclareMathOperator{\Imn}{Im}

\begin{document}
\title{Metal-insulator transition in hydrogenated graphene as
manifestation of quasiparticle spectrum rearrangement of anomalous
type}
 
\author{Yuriy~V.~Skrypnyk}
\affiliation{G. V. Kurdyumov Institute of Metal Physics,\\ 
             National Academy of Sciences of Ukraine,
             Vernadsky Ave. 36,
             Kyiv 03680, Ukraine}
\author{Vadim~M.~Loktev}
\affiliation{Bogolyubov Institute for Theoretical Physics,\\
             National Academy of Sciences of Ukraine,
             Metrolohichna Str. 14-b,
             Kyiv 03680, Ukraine}

\pacs{71.30.+h, 71.23.-k, 71.23.An, 71.55.-i}

\begin{abstract}
We demonstrate that the spectrum rearrangement can be considered as a
precursor and a basis for the metal-insulator transition observed in
graphene dosed with hydrogen atoms. The Anderson-type transition is
attributed to the Fermi level's entering into the quasigap, which
develop in the vicinity of the impurity resonance energy due to the
anomalous spectrum rearrangement. Theoretical results for the Lifshitz
impurity model are in a reasonable agreement with available
experimental data.
\end{abstract}

\maketitle

\section{introduction}

Since the existence of Dirac quasiparticles has been proved for
graphene, one of the most intriguing issues of its physics is the
possibility of their localization by whichever imperfection that can
appear in the honeycomb
lattice.\cite{castr,beena,altsh,gorba,chang,bardas} Moreover, It
should not be overlooked that the effect of disorder on the transport
of quasiparticles is sensitive to the nature of inhomogeneity, be it
caused by short-range or long-range defects, neutral or charged
impurities, adsorbed or interstitial atoms or molecules, vacancies,
spatial distortions or structural irregularities, including ripples on
the graphene sheet and other long-wave random modulations. All these
different types of imperfections necessitate corresponding dedicated
studies, which can not be accomplished by resorting to a single
impurity model. Since these imperfections are rather dissimilar both
in their character and action on basic properties of graphene,
respective theoretical models should be diverse as well.

Early experiments on graphene-based devices, which were engineered
like a commonplace field effect transistors, revealed that the sample
conductivity never drops below a certain \textit{minimal}
value.\cite{natur} This fact, indeed, considerably reduced audacious
expectations that corresponding devices are capable to serve as
next-generation electronic switches. Nevertheless (or, maybe, exactly
because of this), the minimal conductivity existence has produced
quite a stir, and its origin has been relentlessly
debated. Distinctive features of charge carriers, which were shown to
obey the linear dispersion, constituted the core of this
discussion. Uniqueness of the electron subsystem in graphene were
pushed to its limits so much that former physics of semiconductors
were sometimes categorically declared being utterly unsuitable for
this uncommon material. It has been speculated that massless,
according to their Dirac dispersion, charge carriers can not be
localized by any degree of disorder caused by lattice imperfections or
impurity centers. The presumed impossibility to localize Dirac
excitations were directly linked to the minimal conductivity
phenomenon by a simple reasoning. Since the product of the wave vector
modulus and the mean free path is confined from below for propagating
states according to the Ioffe-Regel criterion, the conventional Drude
expression immediately yields the minimum value of $\sim e^{2}/h$ for
each conducting channel. This argument, along with available
experimental results, has led to an opinion that the minimum
conductivity value has a \textit{universal} character for graphene,
and is expressed only through fundamental constants. 

In the mean time, there appeared occasional theoretical studies, which
did not deny a possibility to localize charge carriers in graphene,
and argued for the mobility edge appearance under certain
circumstances, either for a disorder of a general type,\cite{alein,
mirli} or for a specific impurity model, in particular, for
substitutional defects,\cite{naumi} and for the Anderson model of
disorder.\cite{amini,roche,feng} Later, it has become apparent that the
minimal conductivity in graphene does depend on the sample quality and
noticeably varies with the impurity
concentration.\cite{fuhre,mucci,peres,sarma,ugart} Furthermore, not so
long ago, a metal-insulator transition (MIT) in graphene dosed with
atomic, which is essential, hydrogen has been convincingly
observed. The MIT has been manifested by an increase of the room
temperature resistance in about four orders of magnitude.\cite{roten}
The transition has been reported for the grown on SiC surface graphene
with a low hydrogen coverage, namely, around $0.1\%$. The presence of
the mobility gap has been also reported for the fluorinated and ion
bombarded graphene.\cite{savch,ionir} 

As well, corresponding data of ARPES measurements indicated the
disappearance of quasiparticle excitations in the system. The MIT of
the Anderson type comes about when the Fermi level finds itself inside
a domain of \textit{localized} states. Since the graphene in the
actual experiment was pristine before the hydrogen deposition and
possessed sounding metallic properties with conventional Fermi-liquid
behavior, it is tempting to anticipate that some amount of deposited
hydrogen is sufficient to open up a reasonably wide region filled with
localized states, or a \textit{quasigap}, in the spectrum. In essence,
this draft description of the leading to the MIT process corresponds
to the well-known phenomenon of \textit{spectrum
rearrangement},\cite{ivlpo, lifsh} which deals with decisive
modifications in the elementary excitation spectrum upon increasing
the impurity concentration in the system. When the quasigap is located
in the vicinity of the Fermi level position, the MIT can take place in
spite of the rather low hydrogen concentration. Below we are going to
link together the predicted spectrum rearrangement in graphene with
point defects and the observed MIT under the hydrogen
dosing.\cite{roten} The spectrum rearrangement takes place when a
single impurity induces a local or a resonance level in the
spectrum. It has been demonstrated recently that graphene almost
inevitably contains traces of impurity centers, which are capable in
producing resonance states.\cite{nov}  

\section{Criticality and types of spectrum rearrangement}

Let us discuss some general spectral properties of impure systems
and assume for this purpose that inside an isolated host band in a
hypothetical $d$-dimensional system  a single impurity center
accounts for a resonance state with the energy $\varepsilon_{r}>0$,
which is measured against the band edge and has been made
dimensionless by the corresponding bandwidth. The dispersion relation
for this host band is supposed to be of the form
$\varepsilon(\bm{k})\sim k^{p}$. Then, the spatial behavior of the
host Green's function at a given energy inside the band, apart from
any other details, should be governed by oscillations with the
characteristic radius $r\sim\varepsilon^{-1/p}$, which is expressed in
units of the appropriate lattice constant. At the energy
$\varepsilon_{r}$, these oscillations determine the effective radius
$r_{im}\sim\varepsilon_{r}^{-1/p}$ of the impurity state. Through a
closeness of the resonance energy to the band edge, the effective
radius of the impurity state may far exceed the lattice period. As the
impurity concentration is increased, the individual impurity states
are becoming more tightly packed. It can be anticipated that a
significant spatial overlap between these states is capable in
provoking qualitative changes in the spectrum. The average distance
between impurities $\bar{r}\sim c^{-1/d}$, where $c$ is the (relative)
impurity concentration, is gradually shrinking with increasing $c$,
and at some stage of the process becomes of the order of
$r_{im}$. This general condition defines the critical concentration
$c_{SR}\sim\varepsilon_{r}^{d/p}$ of the spectrum rearrangement in the
corresponding impure system. It seems justifiable to reiterate here
that the value of $c_{SR}$ can be fairly low due to the smallness of
the energy $\varepsilon _{r}$.  

Graphene is, evidently, a two-dimensional system, so that $d=2$, and
features the linear dispersion of charge carriers, which implies that
$p=1$. Consequently, the critical concentration of spectrum
rearrangement in the impure graphene is expected to be
$\sim\varepsilon_{r}^{2}$, where energy is counted from the Dirac
point, at which the valence band and the conduction band
coincide. In a case of the conventional dispersion with $p=2$, the
identical in appearance relation between $\varepsilon_{r}$ and
$c_{SR}$ is inherent in four-dimensional systems. In this sense
graphene, as regards the spectrum rearrangement, can be formally
viewed as a conventional system of \textit{increased} spatial
dimensionality. This fact further accentuates the uniqueness of
graphene, since the spectrum rearrangement has not been studied for
such systems.   

Let us return back to experimental data contained in
Ref.~\onlinecite{roten}. The comparison of angle-integrated spectrum
for clean and hydrogen dosed graphene suggests that the dimensional
resonance energy $E_{r}$ lies somewhere around $0.25$ eV lover than
the initial Fermi level, while the Dirac point position $E_{D}$ in the
undoped crystal is about $0.45$ eV below it. A bit more careful
analysis of the spectrum rearrangement in impure graphene,
\cite{skrlo} which is based on the expansion of the self-energy into
a cluster series,\cite{ivlpo} brings about the following expression
for $c_{SR}$, which, for a convenience, is rewritten here in terms of
the hydrogen coverage,
\begin{equation}
n_{SR}\sim n_{C}\frac{(E_{r}-E_{D})^{2}}{W^{2}}, \label{critcov}
\end{equation}
where 
\begin{equation}
n_{C}=\frac{4}{\sqrt{3}a^{2}}\approx3.8\times10^{15}\textrm{cm}^{-2}
\end{equation} 
is the areal density of carbon atoms in graphene, $a=0.246$ nm -- the
lattice constant of graphene, $n_{SR}$ -- the critical coverage of
hydrogen atoms, $W\approx 6.3$ eV -- the bandwidth parameter, and the
logarithmic correction is omitted because of the relatively large
resonance energy. Substitution of the guessed value for the resonance
energy results in $n_{SR}$ that is reasonably close to the reported
MIT critical hydrogen coverage of $3.8\times10^{12}$
cm$^{-2}$.\cite{roten} As a matter of fact, the critical coverage of
the spectrum rearrangement $n_{SR}$, should not be identical to the
experimentally obtained hydrogen coverage required for the MIT, which
will be discussed later.

It should be also noted that there are two main types of the spectrum
rearrangement: the \textit{cross} one, which is usually accompanied by
a sharp single impurity resonance, and the \textit{anomalous} one. The
first type of the spectrum rearrangement results in a quasiparticle
dispersion that looks like a familiar hybridization between the host
branch and a dispersionless branch at the impurity resonance
energy. In this case, two new branches are separated by a gap, which
widens with an increase in the impurity concentration. Consequently,
two different wave vectors correspond to the same energy in this split
spectrum. However, this feature has not been detected in the ARPES
measurements on the hydrogenated graphene. The second type of the
spectrum rearrangement, which is usually encountered in
low-dimensional systems, is of a more diffused nature and frequently
corresponds to a considerably smeared single impurity resonance. This
anomalous spectrum rearrangement is characterized by the opening of
the quasigap, which is filled with localized states and separates two
non-overlapping branches of extended states exhibiting a renormalized
dispersion.  

\section{Interplay between spectrum rearrangement and Anderson
transition}
 
The anomalous type of the spectrum rearrangement has been shown to
unfold in graphene for impurities described by the Lifshitz
model.\cite{skrlo} Within this model, identical impurity centers are
supposed to be randomly distributed on sites of the host lattice. Each
impurity is only allowed to change the on-site energy at its location
in the corresponding tight-binding Hamiltonian. It is well-known
that the electron subsystem in graphene encompasses practically free
$\pi$-electrons, which number is equal to the number of lattice
sites. As soon as an additional hydrogen atom lands on graphene, its
uncoupled s-electron immediately enters a chemical bond with one of
the $\pi$-electrons. That should make the latter one localized almost
completely at the impurity site. In the first approximation, one can
assume that the resulting perturbation comes only from the appearance
of strong attracting potential on the occupied carbon atom. Thus, the
disordered system is described by following operators, 
\begin{equation}
\bm{H}=\bm{H}_{0}+\bm{H}_{im},\qquad\bm{H}_{im}=V_{L}%
\sum\limits_{\bm{n}\alpha}\eta_{\bm{n}\alpha}^{\phantom{\dagger}}
c_{\bm{n}\alpha}^{\dagger}c_{\bm{n}\alpha}^{\phantom{\dagger}},
\label{ham}
\end{equation}
where
\begin{equation}
\bm{H}_{0}=t\sum\limits_{\bm{n}\alpha,\bm{m}\beta}%
c_{\mathbf{n}\alpha}^{\dagger}c_{\mathbf{m}\beta}^{\phantom{\dagger}}
\label{ham0}
\end{equation}
is the host Hamiltonian, in which summation is restricted to nearest
neighbors, $\bm{n}$ runs over lattice cells, indices $\alpha$ and
$\beta$ enumerate sublattices, $c_{\bm{n}\alpha}^{\dagger}$ and
$c_{\bm{n}\alpha}^{\phantom{\dagger}}$ are the creation and
annihilation operators at the corresponding lattice site,
$t\approx2.7$ eV -- is the tight-binding transfer integral for the
$\pi$ bands in graphene, the variable $\eta_{\mathbf{n}\alpha}$ takes
the value of $1$ with the probability $n_{H}/n_{C}$ or the value of
$0$ with the probability $1-n_{H}/n_{C}$, where $n_{H}$ is the
hydrogen coverage, and $V_{L}<0$ is the difference in the
on-site potential at the defect position. 

\begin{figure}
\includegraphics[width=0.96\textwidth]{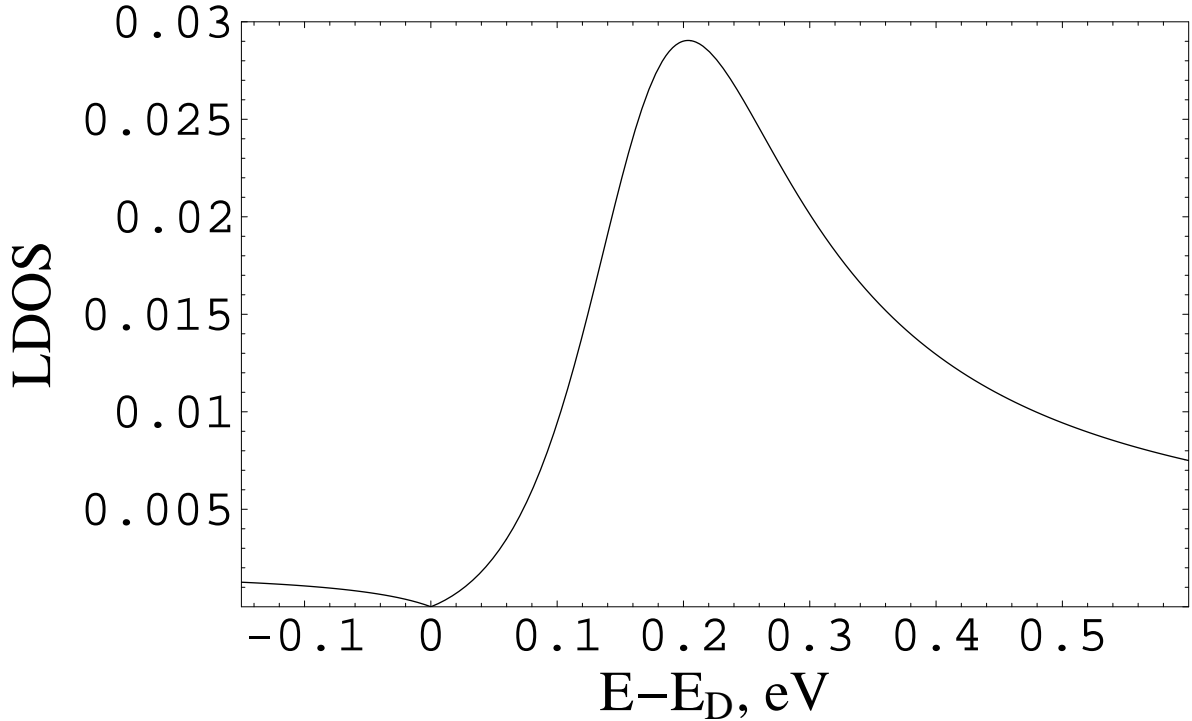}\caption{Local density of
states at the impurity site for $V_{L}=-25$ eV.}
\label{f1}
\end{figure}

Within this model, the local density of states (LDOS) at the lattice
site occupied by a hydrogen reads,  
\begin{equation}
\rho_{im}(E)=-\frac{1}{\pi}\Imn\frac{g_{0}(E)}{1-V_{L}g_{0}(E)},
\label{ldos}
\end{equation}
where $g_{0}(E)$ is the diagonal element of the host Green's function 
\begin{equation} 
\bm{g}=(E-\bm{H}_{0})^{-1}.
\end{equation}
In the vicinity of the Dirac point in the spectrum, namely, within the
window stretching up to around $0.75$ eV to each side of it, this
diagonal element can be easily approximated,
\begin{equation}
g_{0}(E)\approx\frac{E-E_{D}}{W^{2}}\ln%
\left(\frac{E-E_{D}}{W}\right)^{2}-i\pi\frac{|E-E_{D}|}{W^{2}},
\label{apr}
\end{equation}
where 
\begin{equation}
W=\sqrt{\pi\sqrt{3}}t
\end{equation}
is the same bandwidth parameter as in (\ref{critcov}). Substitution of
this approximation to Eq. (\ref{ldos}) yields 
\begin{equation}
\rho_{im}(E)\approx\frac{|\mathcal{E}|}{W[(1-2\frac{V_{L}\mathcal{E}}%
{W}\ln|\mathcal{E}|)^{2}+(\pi\frac{V_{L}\mathcal{E}}{W})^{2}]},
\label{ldosappr}
\end{equation}
where
\begin{equation} 
\mathcal{E}=(E-E_{D})/{W}.
\end{equation} 
The LDOS at the impurity site is shown in Fig. \ref{f1} for
$V_{L}=-25$ eV. As follows from the Figure, the resonance energy
$E_{r}$ is located approximately $0.2$ eV above the Dirac point and
thus corresponds to the experimentally observed peak. While the
required change in the on-site potential is quite substantial, the
large perturbation magnitude should be considered, first of all, as a
result of the chosen simplified impurity model. Unless another is
explicitly specified, the perturbation value of $-25$ eV will be used
below for all subsequent estimations.  

When a certain amount of deposited hydrogen atoms is taken into
account, the negative potentials induced by them on the occupied
lattice sites impede the electron movement. For the Lifshitz impurity
model, which has been described above, the course of the spectrum
rearrangement has been studied both analytically and
numerically.\cite{skrlo, persk} The analysis has shown that states are
subject to localization near the resonance energy, where the impurity
scattering is the strongest. At some coverage, a quasigap occupied by
localized states opens around the energy $E_{r}$. With increasing
impurity concentration, the quasigap gradually broadens and the degree
of localization inside it rises. This quasigap expansion and the
localization enhancement are not symmetric about the energy $E_{r}$
and are more expressed above the resonance (for an attractive
potential). Finally, the quasigap consumes all the space between
$E_{r}$ and $E_{D}$ in the spectrum, which corresponds to the onset of
the spectrum rearrangement by the definition. In other words, the
spectrum rearrangement does follow the anomalous scenario. 

In the experiment discussed, dosed hydrogen atoms act not only as
scattering centers, but also as acceptors. Thus, their presence
inevitably lowers the Fermi level of the system. Therefore, while the
hydrogen coverage is increasing, the Fermi level and the upper
boundary of the quasigap, where the localization of states is most
pronounced, are moving toward each other. Sooner or later, the Fermi
level should appear inside the quasigap, which will cause the MIT of
the Anderson type. Since the energy $E_{r}$ is positioned somewhere
in-between $E_{D}$ and the bare Fermi level $E^{(bare)}_{F}$, the
critical coverage of the spectrum rearrangement should be close but
not identical to the critical coverage of the MIT.

The acceptor effect of the deposited hydrogen atoms is also implicitly
contained in the Lifshitz model. Strong attracting potential $V_{L}<0$
produces a deep local level below the conduction band. Because of its
remoteness from the continuous spectrum, the corresponding narrow
impurity band should hold almost $n_{H}N_{C}/(2n_{C})$ energy levels,
where $N_{C}$ is the total amount of carbon atoms in the system. Thus,
taking into account the spin degeneracy, this impurity band consumes
nearly one electron per deposited hydrogen atom. However, apart from
these qualitative considerations, the adopted impurity model allows to
address the question on the Fermi level position more
precisely. Consider the (normalized) total number of states with
energies that are less than a specified one, 
\begin{equation}
N(E)=\int_{-\infty}^{E}\rho(E)\,\mathrm{d}E,
\end{equation}
where $\rho(E)$ is the density of states. In a disordered system this
quantity can be expanded in powers of the impurity
concentration,\cite{lifsh} 
\begin{equation}
N(E)=N_{0}(E)+\frac{2 n_{H}}{\pi n_{C}}\arg(1-V_{L}g_{0}(E))+\dots,
\end{equation}
where the presence of two sublattices in graphene and the Lifshitz
impurity model are taken into account, and $N_{0}(E)$ is the total
number of states in the host system. Keeping in mind that the deep
impurity band is modeling the acceptor effect of impurities, it is
natural to demand the conservation of the number of occupied states in
the system with varying the amount of dosed hydrogen. This yields a
kind of a balance condition,  
\begin{equation}
\sgn\left(\frac{E_{F}-E_{D}}{W}\right)\frac{(E_{F}-E_{D})^{2}}{W^{2}}
+\frac{2n_{H}}{\pi n_{C}}\arg(1-V_{L}g_{0}(E_{F}))%
=\frac{(E^{(bare)}_{F}-E_{D})^{2}}{W^{2}},
\label{ef}
\end{equation}
where the constant term is dropped from the both sides of the
equation, and Eq.~(\ref{apr}) is used to obtain $N_{0}(E)$ for the
clean graphene. From here and on this condition will be employed to
determine the Fermi level position at a given dosing level. It is
worth mentioning that the Dirac point position (which should not be
confused with its value for the unperturbed system $E_{D}$) is
gradually shifted with increasing the impurity concentration by
approximately $n_{H}V_{L}/n_{C}$. Since the Dirac point and the Fermi
level are both moving in one direction, namely to lower energies, the
distance between them does not shorten remarkably.

\section{Conduction band spectrum}

\begin{figure}
\includegraphics[width=0.96\textwidth]{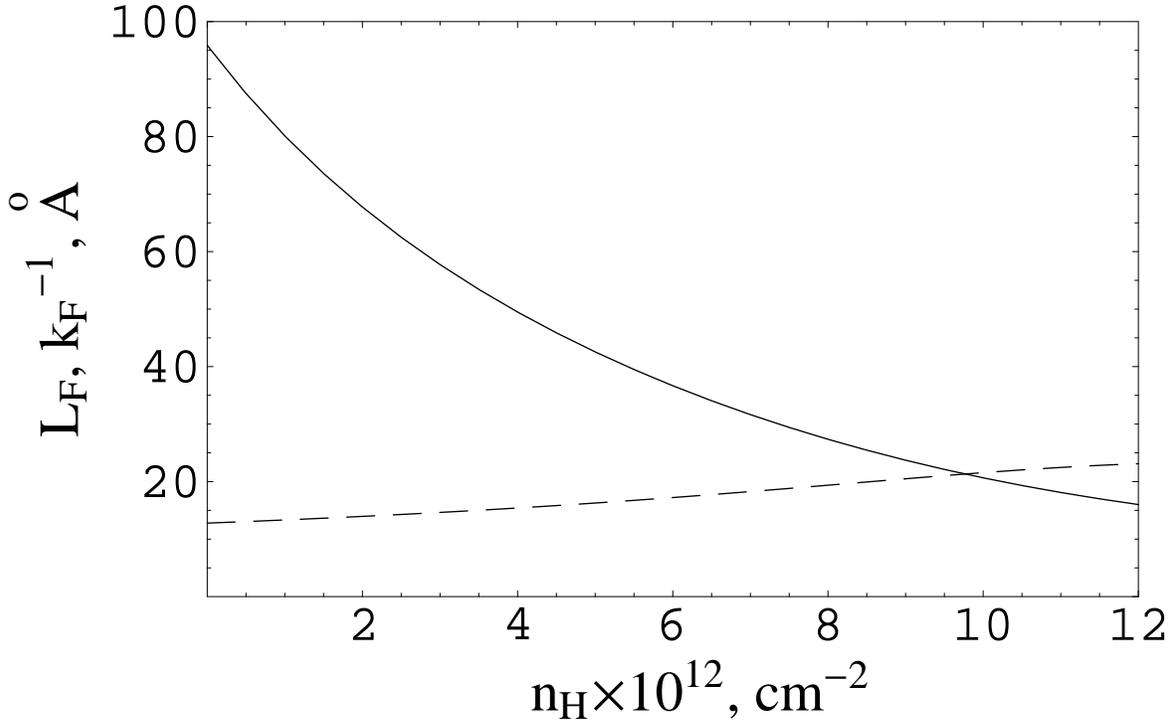}\caption{The  mean free path
$L_{F}$ (solid line) and the inverse wave vector $k_{F}^{-1}$ (dashed
line) at the Fermi level \textit{vs} the hydrogen coverage.}
\label{f2}
\end{figure}

For $n_{H}<n_{SR}$, the presence of attracting impurities does not
distort significantly the quasiparticle dispersion in the valence
band. In contrast, impurities noticeably affect the conduction band. 
The self-energy $\bm{\Sigma}$, which enters the Dyson equation
\begin{equation}
\bm{G}=\bm{g}+\bm{g}\bm{\Sigma}\bm{G}
\end{equation}
for the averaged over impurity distributions Green's function of the
disordered system
\begin{equation}
\bm{G}=\biglb<(E-\bm{H})^{-1}\bigrb>
\end{equation}
can not be attributed only to the impurity scattering in the actual
experiment of Ref.~\onlinecite{roten}, because the linewidth of the
ARPES data is not negligible even at the absence of impurities. To
deal with this issue, we assume that all other sources of its
broadening, except the controlled amount of hydrogen atoms, can be
roughly described by a constant dumping term, 
\begin{equation}
\bm{\Sigma}\approx\bm{\Sigma}_{im}-i\Gamma\bm{I},\qquad \Gamma>0,
\label{sg}
\end{equation}
where $\bm{\Sigma_{im}}$ is those part of the self-energy, which
results from the impurity scattering, and $\bm{I}$ is the identity
matrix. A comparison of the density of states obtained by the
numerical simulation with the one resulted from the standard average
T-matrix method showed that at $n_{H}<n_{SR}$ this approximation
works rather well in the vicinity of the impurity resonance.\cite{persk}
Within this single-site approximation, the impurity-induced part of
the self-energy is diagonal in lattice sites and sublattices for point
defects,
\begin{equation}
\bm{\Sigma}_{im}\approx\frac{\frac{n_{H}}{n_{C}}V_{L}}%
{1-(1-\frac{n_{H}}{n_{C}})V_{L}g_{0}(E)}\bm{I}.
\label{tm}
\end{equation}
Thus, taking into account both assumptions, the total self-energy is
also diagonal,
\begin{equation}
\bm{\Sigma}\approx\Sigma(E)\bm{I}. 
\end{equation}
According to the conventional expression for the spectral function,
\begin{equation}
A(E,\bm{k})\approx-\frac{1}{\pi}\Imn\frac{1}{E-\Sigma(E)-v_{F}k},
\label{sf}
\end{equation}
where
\begin{equation}
v_{F}=\frac{\sqrt{3}at}{2}
\end{equation}
is the Fermi velocity, the wave vector $\bm{k}$ is counted from the
corresponding Dirac point, and only one branch, which is manifested in
the actual experiment, is retained, it is straightforward to put down
corresponding expressions for the inverse width of the momentum
distribution curve at its half-height, 
\begin{equation}
L(E)\approx -\frac{v_{F}}{2\Imn\Sigma(E)},
\end{equation}
and for the inverse of its median
\begin{equation}
k^{-1}(E)\approx\frac{v_{F}}{E-\Sigma(E)}.
\end{equation}
Both this quantities are taken at the Fermi level, $L_{F}\equiv
L(E_{F})$ and $k_{F}^{-1}\equiv 1/k(E_{F})$, and depicted in
Fig.~\ref{f2} in their dependence on the hydrogen coverage. The Fermi
level position is calculated from Eq.~(\ref{ef}), and the external
damping $\Gamma$ is chosen to be $0.03$ eV.  

\begin{figure}
\includegraphics[width=0.96\textwidth]{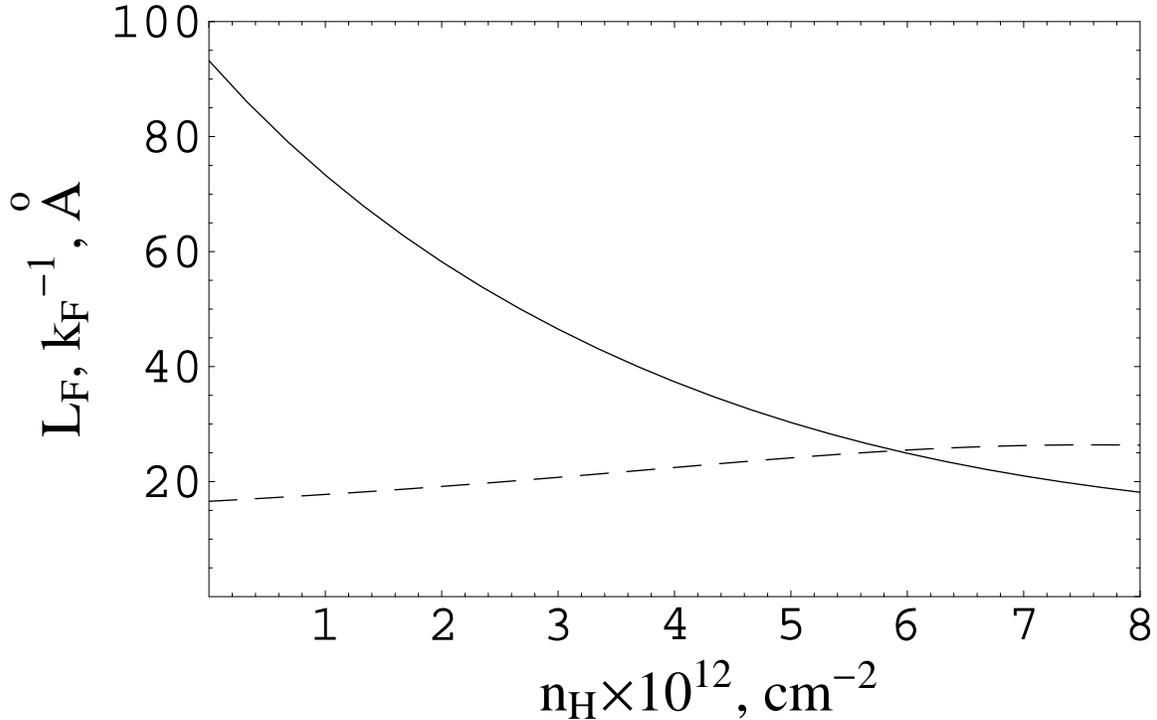}\caption{The  mean free path
$L_{F}$ (solid line) and the inverse wave vector $k_{F}^{-1}$ (dashed
line) at the Fermi level \textit{vs} the hydrogen coverage for
$t=3.5$ eV, $V_{L}=-35$ eV, and $\Gamma=0.04$ eV.}
\label{f3}
\end{figure}

The ARPES linewidth analysis given in Fig.~3(c) of
Ref.~\onlinecite{roten} can be compared with Fig.~\ref{f2}. Since
the inverse momentum width $L(E)$ corresponds to the magnitude of the
mean free path of charge carriers, the Ioffe-Regel
criterion\cite{ioffe} is violated for states at the Fermi level,
\textit{i.e.} $k_{F}L_{F}=1$, at somewhat higher hydrogen coverage,
namely, around $9.5\times 10^{12}$ cm$^{-2}$, than in the experiment
-- $6.5\times 10^{12}$ cm$^{-2}$. This discrepancy, for the most part,
arise from the noticeably reduced wave vectors in the conduction band
of the sample containing no hydrogen. The reduction is evident in
comparison with the dispersion in the valence band, or with the
idealized graphene (\ref{ham0}). The experimentally obtained
quasiparticle dispersion looks already distinctly distorted before any
hydrogen deposition. The distortion can be attributed to the possible
presence of unidentified impurities in the system.\cite{roten} To a
certain extent, this issue can be crudely addressed by a corresponding
adjustment of the hopping integral $t$ in the Hamiltonian. Results of
such quick guesstimate are shown in Fig.~\ref{f3}.

The deliberate adjustment of the transfer integral $t$ pushes
$k_{F}^{-1}$ to where it should be at the zero hydrogen
coverage. Correspondingly, the Ioffe-Regel criterion is violated at
lower hydrogen coverage matching the experimental data. However,
such a forthright measure will be suitable only in the Fermi level
vicinity. It is more consistent to work out a proper model of the host
system that is capable in simulating its experimentally obtained
spectrum. Anyhow, the resulting adjustment of the zero coverage Fermi
wave vector moves the entire curve for the inverse Fermi vector
upwards and, thus, diminish the impurity concentration, at which the
Ioffe-Regel criterion ceases to hold.     

It should be mentioned that, according to the routine of the spectrum
rearrangement analysis, one should expect the Fermi level to enter
inside a domain of localized states at
$k_{F}L_{F}\sim\sqrt{3}/2$. This will happen for the idealized host
spectrum at the hydrogen coverage around $10.5\times 10^{12}$
cm$^{-2}$, which is a little bit higher than the coverage
corresponding to $k_{F}L_{F}=1$. With further increase in the impurity
concentration, the average T-matrix approximation (\ref{tm}) becomes
non-valid because of the increased scattering on impurity clusters,
and the approach outlined above is only suitable for signaling a strong
localization at the Fermi level. On the other side, the experimentally
detected sharp increase in the sample resistance, see Fig.~3(b) of the
Ref.~\onlinecite{roten}, also succeeds the Ioffe-Regel criterion
violation.

\begin{figure}
\includegraphics[width=0.96\textwidth]{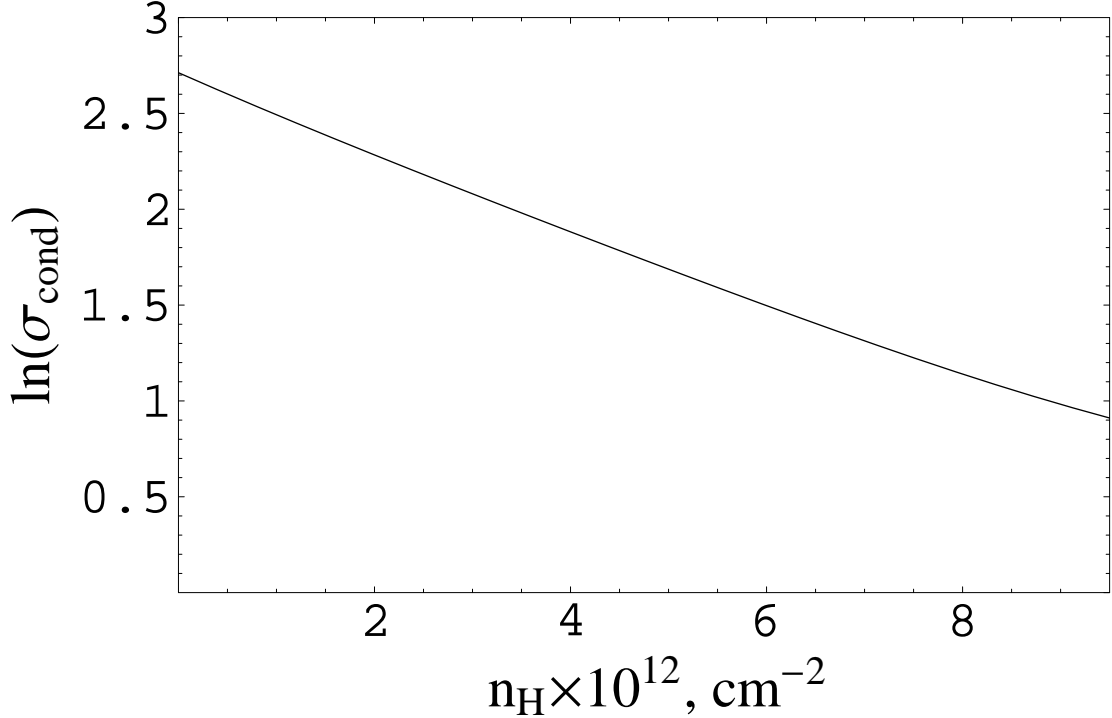}\caption{Logarithm of the
dimensionless zero temperature conductivity \textit{vs} the hydrogen
coverage.} 
\label{f4}
\end{figure}

Within the Kubo approach the zero temperature conductivity can be
written as follows,\cite{minca}
\begin{gather}
\tilde{\sigma}_{cond}%
=\left(\frac{e^{2}}{h}\right)\sigma_{cond}, \\
\sigma_{cond}=\frac{2}{\pi}\left[1+%
(\cot\varphi_{F}+\tan\varphi_{F})\Bigl(\frac{\pi}{2}%
-\varphi_{F}\Bigr)\right], \label{kub}
\end{gather}
where 
\begin{equation}
\varphi_{F}=\arg(E_{F}-\Sigma(E_{F})).
\end{equation}
The dependence of the dimensionless conductivity $\sigma_{cond}$ on
the hydrogen coverage is plotted in Fig.~\ref{f4}. 
It is evident from  Fig.~\ref{f4} that the conductivity is falling
down nearly exponentially with increasing the impurity
concentration. Indeed, the Kubo formula and the average T-matrix
approximation are becoming not so reliable as approaching the mobility
edge. Thus, the zero temperature conductivity can not be
satisfactorily described by Eqs.~(\ref{tm}) and (\ref{kub}) close to
the MIT.

\begin{figure}
\includegraphics[width=0.96\textwidth]{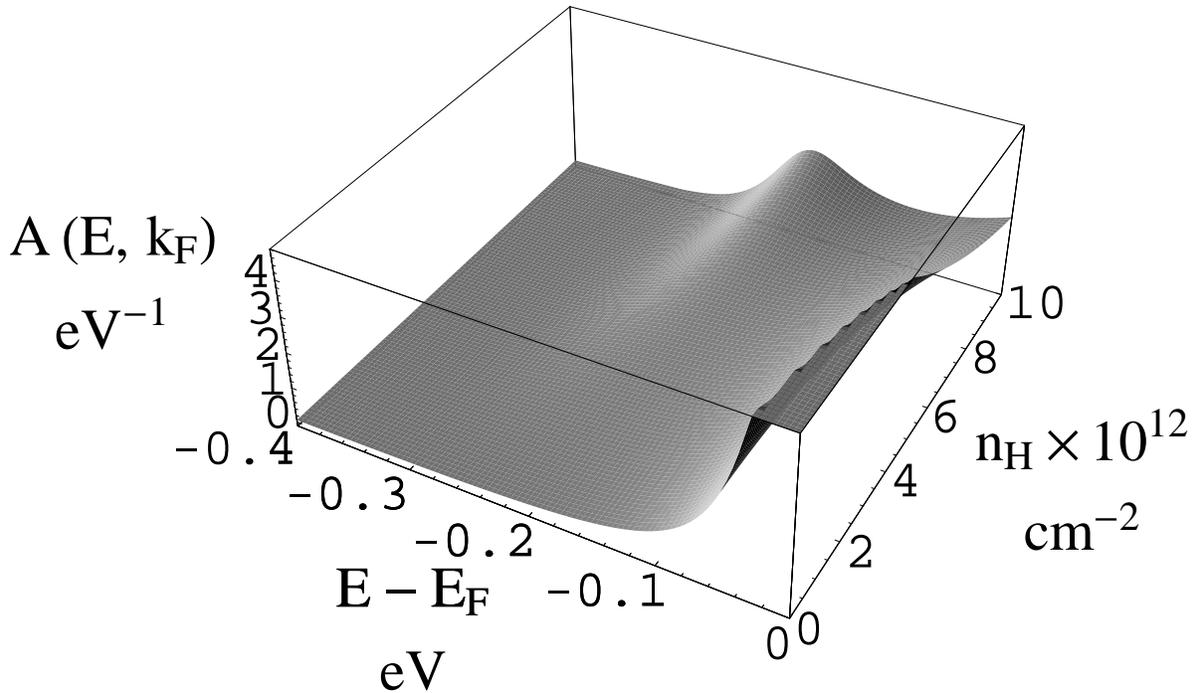}\caption{Evolution of the
energy distribution curve at $k_{F}$ with increasing the hydrogen
coverage.}  
\label{f5}
\end{figure}

The concentration dependence of the energy distribution curve at the
Fermi wave vector $k_{F}$ can be easily re-created for the Lifshitz
impurity model with the help of Eqs.~(\ref{ef}), (\ref{sg}),
(\ref{tm}), and (\ref{sf}). The corresponding magnitude, $A(E,k_{F})$,
is given in Fig.~\ref{f5} as a function of the binding energy
$E-E_{F}$ and the hydrogen coverage $n_{H}$.     
It is clearly seen from the Figure that the energy width of the
quasiparticle peak at $k_{F}$ is gradually increasing with increasing
the hydrogen coverage. Close to the critical impurity concentration
for the MIT, the width of the energy distribution curve occupies
nearly all conduction band below the Fermi level, which indicates the
presence of a developed quasigap in the spectrum. This broadening
translates into the complete breakdown of the quasiparticle picture
near the MIT. The characteristic curve shape consisting of two
competing peaks reproduces well the experimentally observable one,
which can be seen in Fig.~2(c) of Ref.~\onlinecite{roten}. At that,
the peak, which is developing with increasing the hydrogen coverage,
is connected with the impurity resonance. 

\begin{figure}
\includegraphics[width=0.96\textwidth]{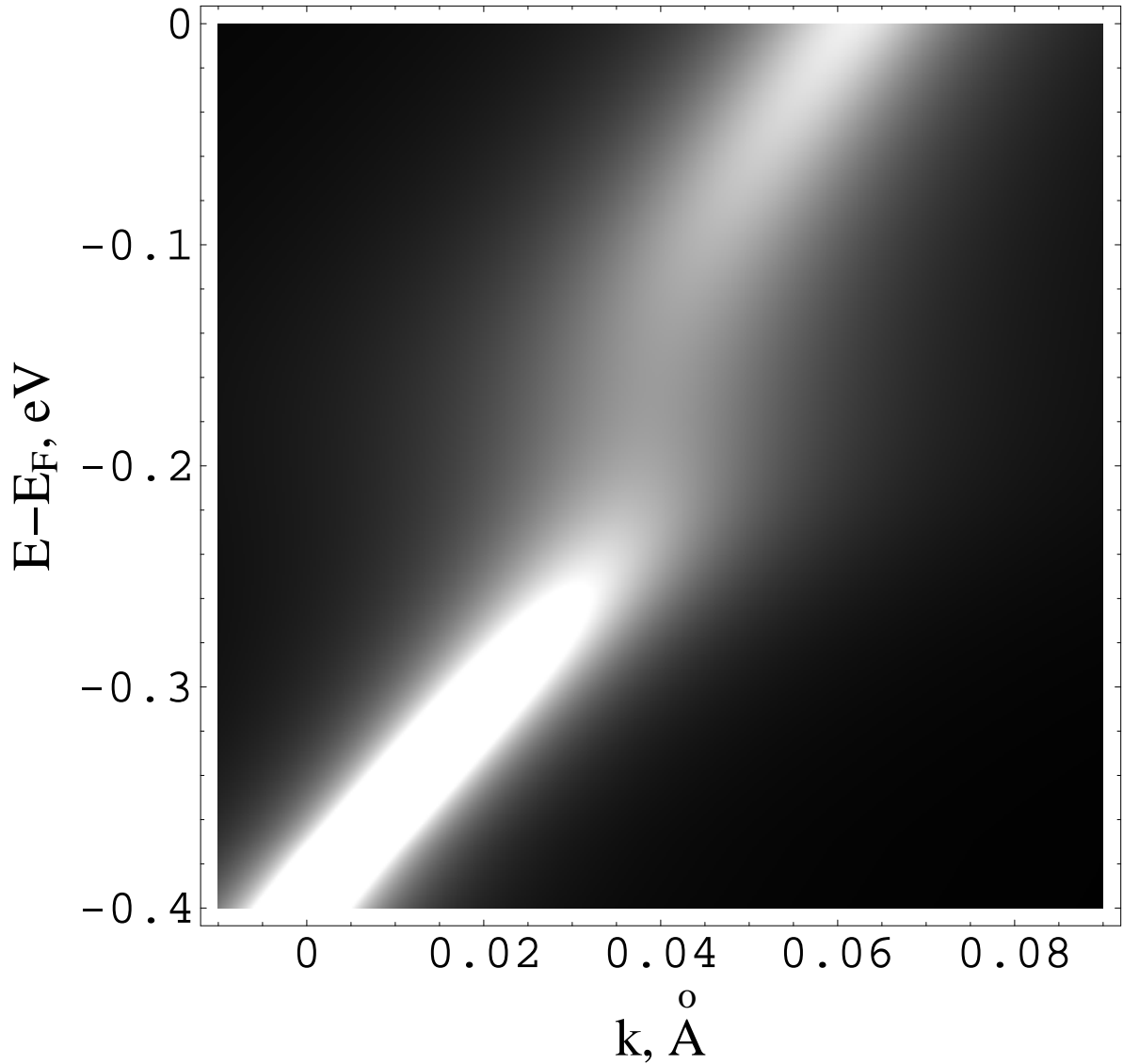}\caption{Density plot of the
spectral function at $n_{H}=5\times 10^{12}$ cm$^{-2}$.}  
\label{f6}
\end{figure}

\begin{figure}
\includegraphics[width=0.96\textwidth]{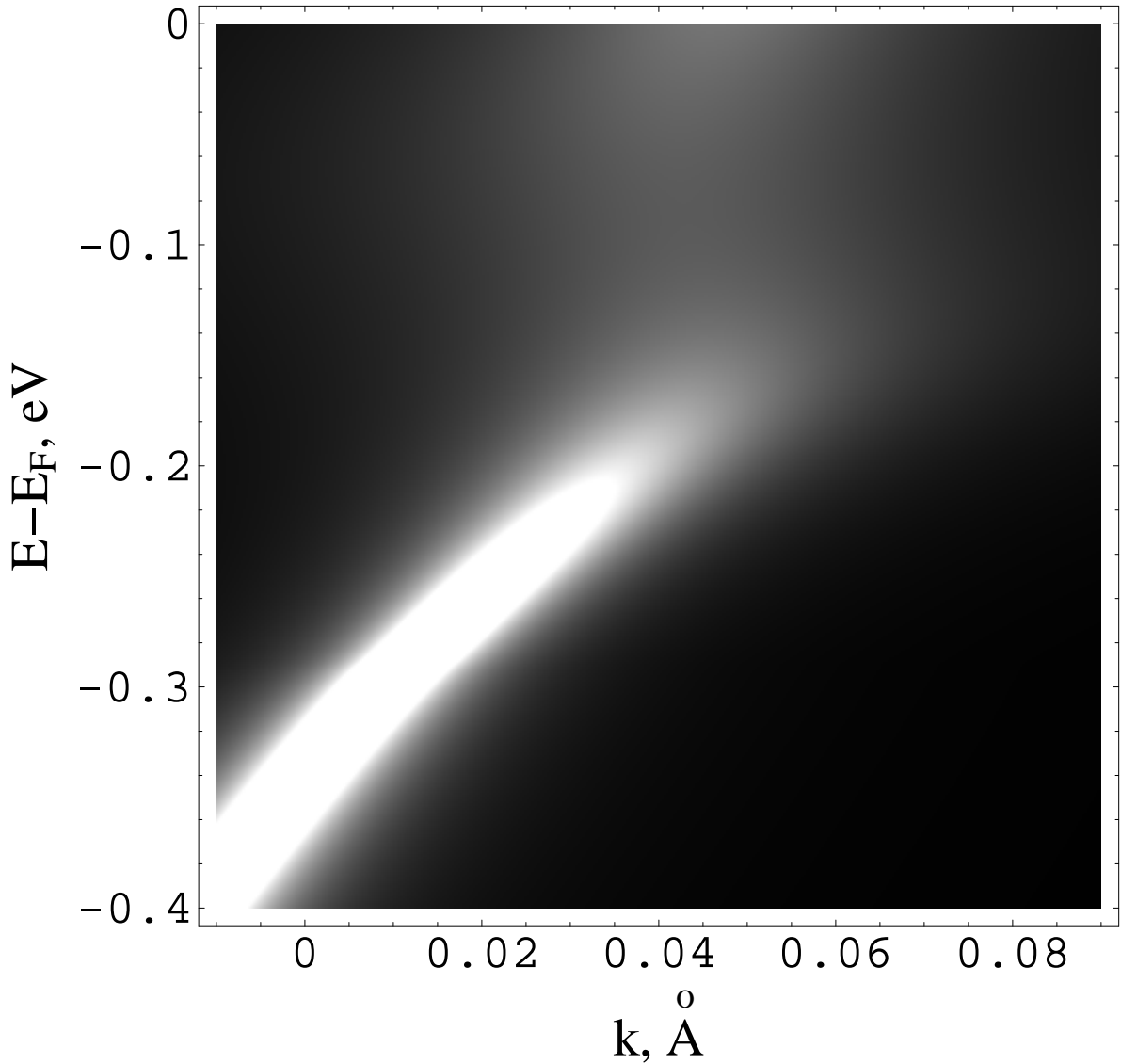}\caption{Density plot of the
spectral function at $n_{H}=10\times 10^{12}$ cm$^{-2}$.}  
\label{f7}
\end{figure}

Finally, we render the density plots of the spectral function
(\ref{sf}) at hydrogen coverages that are close to the critical one
for the MIT in order to reproduce Fig.~1(d) of
Ref.~\onlinecite{roten}. In Fig.~\ref{f6} the Fermi level is about to
enter the spectral region near the resonance energy, where the
quasigap is forming due to strong impurity scattering. In
Fig.~\ref{f7} the Fermi level is already inside the developed
quasigap. The pattern of Fig.~\ref{f7} suggests the apparent upturn of
the dispersion as approaching the Fermi level of the system, which has
been noticed in the experiment. It is worth mentioning here that point
defects does not cause a uniform broadening of the spectral function,
on the contrary, the impurity-induced broadening is most pronounced in
the vicinity of the resonance energy $E_{r}$ in this case.

\section{Conclusion}

To summarize, we have established that point defects are introducing
a new length parameter into the impure graphene. This length parameter
results from the presence of a single-impurity resonance and shows up
as the effective radius of a single-impurity state. When the average
distance between impurities decreases up to this effective radius, the
quasiparticle spectrum undergoes the cardinal rearrangement. The
spectrum rearrangement is manifested in the opening of the quasigap
around the impurity resonance energy in the spectrum. The quasigap
progressively widens with increasing the impurity concentration. The
Fermi level moves due to the doping effect of impurities, as well as
confining the quasigap mobility edges are moving due to its
expansion. If the Fermi level crosses one of the mobility edges and
enters the quasigap, where states are localized, the MIT of the
Anderson type takes place in the disordered system.

We have presented arguments, which confirm that the above scenario can
be considered as a sounding candidate for the explanation of the
experimentally observed MIT in graphen dosed with hydrogen. Even
employing the simple Lifshitz model for the impurity centers, it
appears possible to achieve a semi-quantitative interpretation of the
experimental data. The MIT of the Anderson type in this case is
prompted by the gradual lowering of the Fermi level due to the
acceptor effect of the hydrogen atoms combined with the persistent
raising of the mobility edge due to the ongoing spectrum
rearrangement. Thus, the spectrum rearrangement acts as the
main cause of the MIT, and, as a common phenomenon in semiconductors,
provides the basis for understanding the physics of the
process. Indeed, the oversimplified impurity model is not able to
convey all the detail of the system's behavior. To fulfill this task,
more sophisticated impurity models are required. Among them the
two-parametric s-d model looks like a more natural choice for the
deposited hydrogen atom. However, more sophisticated impurity models
should not change the general physics of the transition process, which
is already captured by the Lifshitz impurity model.   
  
\begin{acknowledgments}
Authors are grateful to Eli Rotenberg for reading the manuscript and
inspiring comments. 

This work was supported by the SCOPES grant No.~IZ73Z0-128026 of Swiss
NSF, the SIMTECH grant No.~246937 of the European FP7 program, the
State Program ``Nanotechnologies and Nanomaterials'', project
No. 1.1.1.3, and by the Program of Fundamental Research of the
Department of Physics and Astronomy of the National Academy of
Sciences of Ukraine.   
\end{acknowledgments}

\end{document}